\documentstyle[12pt]{article}
\let\section=\subsection  \let\subsection=\subsubsection

\def\be{\begin{equation}}
\def\ee{\end{equation}}
\def\bea{\begin{eqnarray}}
\def\eea{\end{eqnarray}}

\begin{document}
\begin{center}
{\large \bf 1-loop correction to the SU(3) symmetric
chiral soliton \footnote{PACS 14.20.-c, 12.40.-y,  Keyword: 
        SU3loop}}\\[5mm]
H.~Walliser \\[5mm]
{\small \it Departamento de F\'{\i}sica, Universidade de Coimbra,\\ 
P-3000 Coimbra, Portugal \\[8mm]} 
\end{center}

\begin{abstract}\noindent
Masses of the SU(3) chiral soliton in tree approximation
turn out at much too high
energies typically around 2 GeV. It is shown that 1-loop corrections
reduce this value drastically with results in the region of the
empirical nucleon mass.
\end{abstract}

\section{Introduction}
In this short letter we discuss the effect of 1-loop corrections to the
mass of the SU(3) symmetric chiral soliton. It is well-known that the
soliton's tree mass turns out too high already in SU(2) and the SU(3)
extension adds further a large kaonic rotational energy such that the
situation is worsened with values which typically lie around or even
above 2 GeV. In SU(2) it was shown \cite{m93,mw96} that pionic 1-loop
corrections are capable to reduce the too high tree mass to a
reasonable number (Table 1). 

The main concern of this paper is to answer the question whether the
same may happen also in SU(3) where we have to start from a much larger
soliton mass, and if that actually occurs how can it be in accordance
with $N_C$ counting? In the course of this investigation we will
recognize as often in soliton models the decisive role of the
Wess-Zumino-Witten (WZW) term.

A detailed discussion of the SU(2) case is found in \cite{mw96} and the
SU(3) extension requires only minor changes in the formulation.
The renormalisation procedure is identical to that used in chiral
perturbation theory and relies on chiral counting together with
dimensional regularisation which preserves chiral symmetry.
Starting point is always an
{\em exact\/} numerical solution of the classical equations of motion 
around which fluctuations for the 1-loop calculation are considered and
the exactness of the solution guarantees that the terms linear in the
fluctuations vanish. These requirements
limit the applicability of the procedure considerably: in SU(3) this is
the rotating hedgehog
\be\label{hedgehog}
U = A U_0 A^{\dagger} \, , \qquad
U_0 = \left( \begin{array}{cc}
e^{i\mbox{\boldmath$ \tau  \hat r$} F(r)} &  \\ & 1 \\
\end{array} \right) \,\, \qquad A \in \mbox{SU(3)}
\ee
in the symmetric case $m_K = m_\pi$. Here $F(r)$ denotes the
chiral angle and $A$ is a SU(3) rotation matrix depending on Euler
angles $\alpha_a \, , \enspace a=1,\dots ,8$.
Already for weak symmetry breaking,
$m_K \stackrel{>}{_\sim} m_\pi$, the ansatz (\ref{hedgehog}) does not
remain exact. Allowing the profile to become Euler angle dependent,
$F(r,\alpha_a)$, improves the situation \cite{sw91} but does not solve
the problem: kaon and eta solitonic components are induced already in
lowest order $(m_K^2-m_\pi^2)$. To consider fluctuations around such a
SU(3) deformed object is certainly beyond present possibilities and we
are therefore not in the position to calculate 1-loop in the
rotator approach \cite{ya88}.
On the other hand, for strong symmetry breaking $m_K \gg m_\pi$ 
the hedgehog (\ref{hedgehog}) rotating in SU(2) only, $A \in $ SU(2),
becomes an exact solution. However there the assumption
$m_K \gg m_\pi$ is in conflict with chiral counting and 
the adopted regularisation scheme. Whether the standard chiral
lagrangian may nevertheless be used in connection with the bound-state
approach \cite{ck88} will be subject of a separate investigation.
At present
our procedure applies only to SU(2) ($m_K \to \infty$) and to SU(3)
symmetry ($m_K = m_\pi$) which is treated in the following. The actual
nucleon mass should lie in between these two limiting cases.

\section{Formulation}

The standard chiral SU(3) lagrangian \cite{gl85} expressed in terms of
the matrix $U$ which contains the dynamical fields and the mass matrix
$M$ 
\bea\label{lagrangian} 
{\cal L} & = & \frac{f^2}{4} tr \left[ \partial_\mu U \partial^\mu U^\dagger
                  + M ( U + U^\dagger ) \right] \nonumber \\
& + &  (L_1 + L_2 + \frac{1}{2} L_3) ( tr \partial_\mu U \partial^\mu
U^\dagger )^2 + \frac{1}{2} L_2 tr([ U^\dagger \partial_\mu U,
U^\dagger \partial_\nu U ] )^2 \nonumber \\
& + &  (L_3 + 3 L_2 ) \left[ tr \partial_\mu U \partial^\mu U^\dagger 
\partial_\nu U \partial^\nu U^\dagger - \frac{1}{2} 
( tr \partial_\mu U \partial^\mu U^\dagger )^2 \right] \nonumber \\
& + & L_4 tr M ( U + U^\dagger) tr \partial_\mu U \partial^\mu U^\dagger
 +  L_5 tr (U M + M U^\dagger )  \partial_\mu U \partial^\mu U^\dagger
\nonumber \\
& + &  L_6 \left( tr M ( U + U^\dagger) \right)^2 
+ L_7 \left( tr M ( U - U^\dagger ) \right)^2
\nonumber \\
& + & L_8 tr ( M  U M U+ M U^\dagger M U^\dagger ) \nonumber \\
& \equiv & \frac{f^2}{4} tr \left[ \partial_\mu U \partial^\mu U^\dagger
                             + M ( U + U^\dagger ) \right]
+ \sum_{i=1}^8 L_i {\cal L}_i^{(4)}
\eea 
comprises the familiar non-linear sigma (N$\ell \sigma$) model of
(chiral order) ChO2 and eight terms of ChO4 which are relevant in the
soliton sector without external fields. At scale 
$\mu = m_\varrho = 770$ MeV
which should provide the lagrangian in leading order $N_C$ \cite{e95}
the renormalized low energy constants (LECs) are chosen
\bea
&& L_1 + L_2 + \frac{1}{2} L_3 = 0 \, , \qquad L_3+3 L_2 = 0 \, ,\qquad
L_2 =  \frac{1}{16e^2} \nonumber \\
&& L_4 = L_6 = 0  \, , \qquad L_5 = 2L_8 = -6L_7 =
\frac{f_K^2 - f_\pi^2}{8(m_K^2 - m_\pi^2)} = 2.3 \cdot 10^{-3}  
\eea 
to be in accordance with the standard values \cite{gl85,e95}
(within error bars) with one exception: $L_2$ has to be fixed by an
effective Skyrme parameter $e=4.25$ (the standard value would
correspond to $e \simeq 7$) in order to simulate the missing higher
ChOs generated by vector mesons. A detailed justification of this
choice is found in ref.  \cite{mw96}. The SU(2) reduction of
(\ref{lagrangian})  yields exactly the lagrangian employed in that
reference (and the LECs which additionally appear in SU(3) take their
standard values). As a consequence the soliton as well as the pionic
1-loop results do not have to be recalculated, but
may just be taken from there. It should be mentioned that although
the LECs are chosen such that many of the ChO4 terms in
(\ref{lagrangian}) vanish at scale $\mu = m_\varrho$ all these terms are
switched on and do contribute when the scale is changed.

In the SU(3) symmetric case under consideration the mass matrix 
$M = m^2 \cdot${\bf 1} is diagonal and leads to
identical kaon and pion masses and decay constants
\bea
&&f_K^2 = f_\pi^2 = f^2 + 8(3L_4+L_5) m^2 \qquad \qquad \qquad \qquad
f = 91.1 \, \mbox{MeV} \nonumber \\
&&f_K^2 m_K^2 = f_\pi^2 m_\pi^2 = f^2 m^2 + 16(3L_6+L_8) m^4 
\qquad \quad m=138 \, \mbox{MeV}
\, .  
\eea 
Because the
symmetry breakers are absent, the nucleon mass in tree approximation
\bea \label{tree}
E_{\mbox{\scriptsize\ tree}}
&=& M_0 + \frac{J(J+1)}{2 \Theta_\pi} + \frac{1}{2 \Theta_K}
\left[ C_2 - J(J+1) - \frac{N_C^2}{12} \right] 
\nonumber \\
&=& M_0 + \frac{3}{8 \Theta_\pi} + \frac{N_C}{4 \Theta_K} \, ,
\qquad N_C \enspace \mbox{odd}
\eea
comprises the soliton mass $M_0$ of order $N_C$, the pionic rotational
energy of order $N_C^{-1}$ ($\Theta_\pi$ pionic moment of inertia) and
the kaonic rotational energy of order $N_C^0$ 
($\Theta_K$ kaonic moment of inertia). The non-trivial $N_C$ assignment
to the kaonic rotational energy is caused by the WZW term which selects
the lowest lying multiplet depending on the number of colors. For odd
$N_C$ the "nucleon" with spin and isospin $1/2$ and hypercharge $N_C/3$
sits in the multiplet with the labels
\be
(p,q)=(1,\frac{N_C-1}{2}) \, , \qquad \qquad
C_2=\frac{N_C^2}{12} + \frac{N_C}{2} + \frac{3}{4} \, .
\ee
With that eigenvalue $C_2$ of the Casimir operator eq. (\ref{tree})
is immediately verified.

For the 1-loop calculation fluctuations $\eta_a$ 
are introduced through the ansatz
\be
\label{ansatz}
U = A \sqrt{U_0} e^{i \lambda_a \eta_a /f} \sqrt{U_0} A^\dagger \, ,
\qquad a=1,\dots,8
\ee
and the corresponding equations of motion (e.o.m.) which according 
to their time dependence $\sim e^{-i\omega t}$ may be written as 
\be \label{eom}
h^2_{ab} \eta_b = \omega^2 n^2_{ab} \eta_b 
\ee
($h^2_{ab}$ is a differential operator and $n^2_{ab}$ the metric)
have to be solved for the phase-shifts. Because the e.o.m. (\ref{eom})
decouple for the different meson species into partial waves characterized
by phonon spin $L$ and parity the pionic, kaonic and eta phase-shifts
may be summed up separately over the various channels $(L c)$
\be
\delta^x(p)=\sum_{Lc} (2L+1) \delta^x_{Lc}(p) \, \qquad
x = \pi , K , \eta \, .
\ee
The ultra-violet divergencies contained in the Casimir energy
are related to the high momentum behaviour of these phaseshifts
\be
\label{deltas}
\delta^x (p) \stackrel{p \to \infty}{\longrightarrow} a^x_0 p^3 + a^x_1 p +
\frac{a^x_2}{p} + \, \cdots
\ee
with expansion coefficients $a^x_0, a^x_1, a^x_2$ known analytically
for the N$\ell \sigma$ model
(the explicitly denoted terms give rise to at least logarithmically
divergent expressions). These coefficients obey the important ChO4
relation
\be
\sum_x \left[ 3 \pi m_x^4 a^x_0 - 4\pi m^2_x a^x_1 + 8 \pi a^x_2 \right]
= \sum^8_{i=1} \Gamma_i \int \, d^3 r {\cal L}_i^{(4)} \, ,
\ee
where the $\Gamma_i$'s are simple
numerical factors given in \cite{gl85} and which is used below for
regularisation of the Casimir energy.
For the full model (\ref{lagrangian}) the
coefficients have to be determined numerically and the challenge is to
calculate the phase-shifts with great precision up to $p_{max} \simeq
25 m_\pi$ where $L_{max} \simeq 100$ partial waves are needed (for
details see \cite{mw96}). With these informations at hand the
divergencies in the 1-loop contribution may be isolated using
dimensional regularisation  
\bea
\label{casimir}
E_{\mbox{\scriptsize\ cas}} 
& = & \frac{1}{2 \pi} \sum_x \left\lgroup - \int_0^{\infty}
\frac{p dp}{\sqrt{p^2 + m^2_x}} [\delta^x (p) - a^x_0 p^3 - a^x_1 p - 
\frac{a^x_2}{p}]  - m_x \delta^x (0)  \right. \nonumber\\
&& \qquad + \left. \frac{3 m^4_x a^x_0}{16} (\frac{1}{6} 
+ \ell n \frac{m^2_x}{\mu^2})
 - \frac{m^2_x a^x_1}{4} \ell n \frac{m^2_x}{\mu^2} + \frac{a^x_2}{2}
(1 + \ell n \frac{m^2_x}{\mu^2})  \right\rgroup  \nonumber \\
&& + \Lambda (\mu) \sum_x \left[ 3 \pi m_x^4 a^x_0 - 4\pi m^2_x a^x_1 + 
8 \pi a^x_2 \right] \\ 
& \equiv & \sum_x E^x_{\mbox{\scriptsize\ cas}} (\mu) + \Lambda (\mu) 
\left[ \sum^8_{i=1} \Gamma_i
\int \, d^3 r {\cal L}_i^{(4)} + \mbox{higher ChOs} \right] \, , \nonumber 
\eea
which involves a scale $\mu$ to render the arguments in the logarithms
dimensionless. The divergencies as $d \to 4$ reside in
\be \label{lambda}
\Lambda (\mu) = \frac{\mu^{d-4}}{16 \pi^2} \left[ \frac{1}{d-4} - \frac{1}{2}
(\Gamma'(1) + \ell n (4 \pi) + 1) \right]
\ee
and may finally be absorbed into a redefinition of the LECs
\be \label{lec}
L^r_i (\mu) = L_i - \Gamma_i \Lambda (\mu) \, , \qquad
L^r_i(\mu)=L_i^r(m_{\varrho})-\frac{\Gamma_i}{32\pi^2}
\ell n(\frac{\mu^2}{m_{\varrho}^2}) 
\ee
which become scale-dependent. The renormalisation scheme
is identical to that used in chiral perturbation theory. 

From (\ref{casimir}) it is also noticed that the regularisation
scheme must fail for $m_K \gg m_\pi$: in the limit $m_K \to \infty$
we would obtain an infinite contribution from the second row
containing the chiral logarithms (the term $m_K \delta^K (0)/2\pi$
would cancel the bound state contribution $\frac{1}{2} \sum_z \omega_z$
which has to be added in that case because the infinitesimal kaonic
rotations appear at finite energies).

For $m_K = m_\pi$ with the finite contributions in (\ref{casimir}) 
the nucleon mass in
tree + 1-loop is finally determined 
\be \label{loop}
E_{\mbox{\scriptsize\ tree+1-loop}} = M_0(\mu) + \frac{3}{8 \Theta_\pi(\mu)} 
+ \frac{N_C}{4 \Theta_K(\mu)} + \sum_x E^x_{\mbox{\scriptsize\ cas}}(\mu) \, . 
\ee
All quantities involved become scale-dependent in a non-trivial way,
this will be investigated. Because the contributions of the various
mesons enter additively we may consider them separately.

\subsection{Pions}

The pionic contribution is the same as in SU(2),
this was mentioned already.
In ref. \cite{mw96} we obtained a 1-loop contribution 
$E^\pi_{\mbox{\scriptsize\ cas}}(m_\varrho)=-680$ MeV at scale $\mu = m_\varrho$. Further it
was found that the scale-dependences of $M_0(\mu)$ and
$E^\pi_{\mbox{\scriptsize\ cas}}(\mu)$ cancel almost exactly over a wide region of scales,
compare Fig. 3.2 in that reference. This finding was interpreted as
strong evidence for the reliability of the renormalisation procedure
and also for the reasonable choice of the effective Skyrme parameter
$e$. The pionic rotational energy in (\ref{tree}) is very small by
itself and its scale-dependence is relatively weak such that it
does not destroy this property.

\subsection{Eta}

The coupling of the $\eta$ to the soliton proceeds through the mass terms
only and consequently is extremely weak. The resulting 1-loop
contribution $E^\eta_{\mbox{\scriptsize\ cas}}(m_\varrho) = +0.5$ MeV is tiny.

\subsection{Kaons}

Because in the kaonic sector we have four infinitesimal kaonic rotations
the phase-shift starts at $\delta^K(0) = 4\pi$ according to Levinson's
theorem.
\begin{figure}[h]
\vspace*{6cm}
\begin{center} \parbox{10cm}{\caption{Scale-dependence
of the kaonic contributions without WZW in tree (dashed) and tree + 1-loop 
(solid). The inclusion of the 1-loop contribution amplifies the
scale-dependence.}} 
\end{center}
\end{figure}
This fact, although the zero-modes do not contribute by themselves
(because they are located at zero energy), 
leads to a large negative Casimir energy which at scale $\mu = m_\varrho$
amounts to $E^K_{\mbox{\scriptsize\ cas}}(m_\varrho) = -425$ MeV. This contribution
compensates nicely for the kaonic rotational energy 
$N_C/4\Theta_K(m_\varrho)=+390$ MeV which is of the same order $N_C^0$
(Table 1). However, the scale-dependence of the rotational energy is
enhanced further if the 1-loop contribution is included as is noticed
from Fig.1. This indicates that there is an important term missing.

\subsection{Kaons with WZW term}

Inclusion of the WZW term has two effects (i) it adds a
contribution to the kaonic moment of inertia and (ii) it modifies the
e.o.m. for the fluctuations (\ref{eom}) introducing a term linar in
$\omega$.
\begin{itemize}
\item[(i)] Kaonic moment of inertia

The WZW term provides the driving term for an induced soliton component
proportional to the kaonic angular velocity which leads to a contribution
to the moment of inertia \cite{wspm90}. Because we have already taken into
account the collective rotation and in order to avoid double counting
we have to impose the constraint
\be \label{constr}
\int d^3\!r\,\, z_a^e n_{ab}^2 \eta_b=0 \, , \qquad 
z_a^e=\frac{2f_K}{\sqrt{\Theta_K}} \sin(\frac{F}{2})
\; f_{aei} \hat r_i 
\ee
that the induced component $\eta$ be orthogonal to the infinitesimal
rotation (for $m_K = m_{\pi}$ the corresponding equation without
constraint would not even have a unique solution because the zero-mode
as solution of the homogeneous equation can always be added with
arbitrary strength \cite{w92}). The constraint reduces the induced
component considerably. For $m_K = m_{\pi}$ and scale $\mu=m_\varrho$
we obtain $\Theta_K(m_\varrho)=(1.92+0.21) \, \mbox{GeV}^{-1} =
2.13 \, \mbox{GeV}^{-1}$ a $10\%$ contribution from the induced
component which however decreases
rapidly with increasing kaon mass justifying that this
contribution is normally neglected. In our case it helps to soften the
scale-dependence of the kaonic rotational energy as is noticed by
comparing Fig.1 with Fig.2.
\item[(ii)] Phase-shifts

The WZW term introduces a term linear in $\omega$ into the e.o.m. and
as a consequence not all infinitesimal rotations remain at zero energy.
For physical kaon mass, two of them appear as bound states and are
interpreted as kaonic excitations of the nucleon (bound-state approach
\cite{ck88}). For $m_K = m_{\pi}$ these states appear as
resonances in the continuum and consequently the kaonic phase-shift
starts only at $\delta^K(0)=2\pi$. Again, because these states are already
considered in our approach in the kaonic collective rotation they have
to be projected from the space of allowed fluctuations by implementing
the constraint (\ref{constr}). Then the phase-shift starts again at
$\delta^K(0)=4\pi$ as it should.
\end{itemize}
\begin{figure}[h]
\vspace*{6cm}
\begin{center} \parbox{10cm}{\caption{Same as Fig.1
but with the WZW term considered. The tree + 1-loop contribution
is almost scale-independent over a wide region of scales}} 
\end{center}
\end{figure}
With inclusion of the WZW term the Casimir energy $E^K_{\mbox{\scriptsize\ cas}}(m_\varrho)
= -600$ MeV at scale $\mu=m_\varrho$ over-compensates the kaonic 
rotational energy $N_C/4\Theta_K(m_\varrho)=+350$ MeV (Table 1).
In contrast to the case without WZW the scale-dependences of these
quantities are opposite such that the kaonic energy in tree +
1-loop becomes almost scale-independent over a wide region (Fig.2)
quite similar to the pionic contributions (for very small scales
the soliton becomes unstable due to a too strong symmetric ChO4 term).
This confirms that the WZW term plays an important role also in this
context.

\section{Results}

Table 1 comprises the results for the nucleon mass in the SU(3) symmetric
chiral soliton model with and without the WZW term taken into account.
It should be kept in mind that there are no additional adjustable
parameters in the game, the effective Skyrme parameter was taken as in 
the SU(2) calculation. It is noticed that in both cases the tree mass of
$\simeq 2$ GeV is appreciably reduced into the region of the physical nucleon
mass. This is compatible with $N_C$ counting because for the kaonic
contributions tree and 1-loop are of the same order $N_C^0$ and because
a strong cancellation between the two occurs. 
\begin{table}[h]
\begin{center}
\parbox{12cm}{\caption{\label{su3} 
Nucleon mass in SU(2) and symmetric SU(3) with and without WZW term.
The individual tree and 1-loop
contributions
are listed separately and their $N_C$ order is indicated.
All energies are given in MeV.}} 
\begin{tabular}{|cc|r|r|r|}
\hline
                &            & SU(2) &   SU(3)$\qquad$& SU(3)$\quad$ \\       
                &            &       &   without WZW & with WZW \\
\hline 
soliton mass    & $N_C^1$    & $1630$  &  $1630 \qquad$  & $1630 \quad$     \\
pionic rotation & $N_C^{-1}$ &  $ 70$  &    $70 \qquad$  &   $70 \quad$     \\ 
kaonic rotation & $N_C^0$    & -$\enspace$ &   $390 \qquad$  &  $350 \quad$ \\ 
\hline
total tree      &            & $1700$  &  $2090 \qquad$  & $2050 \quad$     \\
\hline
$\pi$ 1-loop    & $N_C^0$    & $-680$   & $-680 \qquad$ & $-680 \quad$    \\
$K$ 1-loop      & $N_C^0$    & -$\enspace$ &  $-425 \qquad$ & $-600 \quad$  \\
$\eta$ 1-loop   & $N_C^0$    & -$\enspace$ &  $+0.5 \qquad$ & $+0.5 \quad$  \\
\hline
total tree + 1-loop &        & $1020$   &   $985 \qquad$ &  $770 \quad$      \\
\hline
\end{tabular}
\end{center} 
\end{table}
As was discussed
in the previous section, scale-independence requires the inclusion of
the WZW term. Thus, the results for the nucleon mass are in case of SU(3)
symmetry $770$ MeV and for SU(2) $1020$ MeV such that the 
empirical nucleon mass
actually lies in between these two limiting values. 
Once the lagrangian is fixed the calculation presented is exact 
to order $N_C^0$, there are no other contributions to this order. 
Unfortunately we
were not in the position to calculate 1-loop corrections for finite
$m_K \neq m_\pi$ for the reasons discussed in the introduction.
Possibly one can estimate the Casimir energy for
$m_K \stackrel{>}{_\sim} m_\pi$ in the rotator approach where one finds
Euler angle dependent scattering equations \cite{s92}, but it is
difficult to control the necessary approximations (neglection of the
terms linear in the fluctutions which arise from the rotating hedgehog
being not an exact solution). On the other hand for $m_K \gg m_\pi$
the 1-loop contributions could possibly be calculated using the
bound-state approach \cite{ck88} which also remains to be
investigated.

\subsection*{Acknowledgements}
This work is supported by JNICT, Portugal (Contract PRAXIS/2/2.1/FIS/451/94).
The author would like to thank G. Holzwarth and N. Scoccola
for helpful discussions.


\begin{thebibliography}{80}
\bibitem{m93}     B. Moussallam, Ann. Phys. {\bf 225} (1993) 264
\bibitem{mw96}    F. Meier and H. Walliser, hep-ph/9602359, 
                  will be published in Phys. Rep. 
\bibitem{sw91}    B. Schwesinger and H. Weigel, Phys. Lett. {\bf B267} 
                  (1991) 438
\bibitem{ya88}    H. Yabu and K. Ando, Nucl. Phys. {\bf B301} (1988) 601
\bibitem{ck88}    C.G. Callan, K Hornbostel and I. Klebanov, 
                  Phys. Lett. {\bf B202} (1988) 269
\bibitem{gl85}    J. Gasser and H. Leutwyler, Nucl. Phys. {\bf B250} (1985) 465
\bibitem{e95}     G. Ecker, Czech. J. of Phys. {\bf 44} (1995) 405
\bibitem{wspm90}  H. Weigel, J. Schechter, N.W. Park and U.G. Meissner,
                  Phys. Rev. {\bf D42} (1990) 3177
\bibitem{w92}     H. Walliser, Siegen preprint 1992, unpublished
\bibitem{s92}     B. Schwesinger, Nucl. Phys. {\bf A537} (1992) 253
\end{thebibliography}
\end{document}